\RequirePackage{fix-cm}
\documentclass[smallextended]{svjour3}       
\smartqed  
\usepackage{url}
\usepackage{graphicx}
\begin{document}

\title{Separability and entanglement in the Hilbert space reference frames related through the generic unitary transform for four level system}


\author{V.I. Man'ko       \and L.A. Markovich}


\institute{V.I. Man'ko \at P. N. Lebedev Physical Institute, Russian Academy of Sciences\\
Leninskii Prospect 53, Moscow 119991, Russia\\
Moscow Institute of Physics and Technology\\
 Institutskii Per. 9, Dolgoprudny Moscow Region 141700, Russia\\
              \email{manko@lebedev.ru}           
           \and
           L.A. Markovich \at
          Moscow Institute of Physics and Technology\\
 Institutskii Per. 9, Dolgoprudny Moscow Region 141700, Russia\\
Institute for information transmission problems, Moscow,\\ Bolshoy Karetny per. 19, build. 1, Moscow 127051, Russia\\
V. A. Trapeznikov Institute of Control Sciences, Moscow,\\Profsoyuznaya 65, 117997 Moscow, Russia\\
              \email{kimo1@mail.ru}
}

\date{Received: date / Accepted: date}

\maketitle

\begin{abstract}
Quantum correlations in the state of four-level atom are investigated by using generic unitary transforms of the classical (diagonal) density matrix. Partial cases of pure state, $X$-state, Werner state are studied in details. The geometrical meaning of  unitary Hilbert reference-frame rotations generating entanglement in the initially separable state is discussed. Characteristics of the entanglement in terms of concurrence, entropy and negativity are obtained as functions of the unitary matrix rotating the reference frame.
\end{abstract}

\section{Introduction}
\label{intro}
Notions of entanglement and separability are central in quantum information processing and computation \cite{Bennett,DiVincenzo}.
The entanglement exists due to quantum correlations between physical systems and it cannot be explained in terms of correlations between
local classical properties inherent in subsystems. The existence of such correlations were stated in 1935 by Einstein, Podolsky and Rosen (EPR) in their famous paper \cite{Einstein} and the term "entanglement" itself was proposed by Schr\"{o}dinger as a reaction to the EPR contribution \cite{schredinger:35}.
It is often said, that the entangled composite system state is nonseparable. The strict definition is that the state
of the composite system  is called separable if it can be written as a decomposition in terms of product states.
The separable state has no quantum entanglement and the nonseparable state is entangled.
\\ The notion of the entanglement is not only theoretical. In the end of the last century it was realized that the quantum entanglement can be used in practice.
Algorithms based on laws of the quantum mechanics were used to solve hard tasks faster then  any known classical computer. Famous examples are the Deutsch's algorithm \cite{Deutsch}, the factorization of the prime numbers by Shor's algorithm \cite{Shor} and the search Grover's algorithm \cite{Grover}. Moreover, the quantum systems have potential in other various kinds of applications and the entanglement plays a key role \cite{Feynman}. Also the  entanglement notion is central in many applications of the quantum information like  quantum teleportation or quantum error correction \cite{Lo,Bennett,Shor,Steane}. It is widely used in quantum key distribution \cite{Curty,Curty2,Ekert}. The entangled states of more then two subsystems were generated in several set ups in the laboratories \cite{Pan,Pan2}. Recently, the entanglement was used for deep-space communications and cryptography, e.g.,
 NASA's Lunar Atmosphere Dust and Environment Explorer decided to use the quantum entanglement for uploading and downloading information
 between the spacecraft and a ground-based receiver \cite{LADEE}. Furthermore, several works appeared where the entanglement is considered as a crucial phenomenon
in biological and chemical systems, e.g., see \cite{Rieper}. That is why their is no doubts that finding the states entanglement detection method is a crucial task in the quantum information.
\subsection{Contributions of this paper}
\par  It is known, that the entanglement does not directly depend on spectrum of the state. In other words, solving the entanglement problem is not the solving the density-operator spectral problem. Different states may have the same spectrum, but one can be entangled, and the second one can be separable.
In literature, there are several methods for detecting the entanglement \cite{Vidal,Hill:1997}. One of them is the Positive Partial Transpose (PPT)  introduced by Peres \cite{Peres}
and Horodecki \cite{Horodecki} as a necessary condition for the joint density matrix of two quantum systems to be separable. Firstly, the PPT was proposed as a necessary condition for every separable state by Peres in 1996. Hence, every separable state must satisfy this condition, but some entangled states can satisfy it, too.
After that, Horodecki proved that the PPT is a necessary and a sufficient condition for separable states of $2\times2$ and $2\times3$ dimensions.
The PPT of the composite density matrix is given by transposing only one of the subsystems of the quantum system.
However, there are several other methods of the entanglement detection like negativity and concurrence \cite{Zyczkowski2,Horodecki2}.
The idea of the negativity is that the entanglement of the quantum system depends on the sum of  modulus eigenvalues of the PPT density matrix of the state.
The concurrence is based on the square roots of the eigenvalues of the special Hermitian matrix formed from the density matrix of the state.
\par  Recently, it was observed in \cite{Mancini,Chernega,Chernega:14,OlgaMankoarxiv} that the quantum properties of the systems without subsystems (a single qudit) can be formulated by using an invertible map of integers $1,2,3\ldots$ onto the pairs (triples, etc.) of integers $(i,k)$, $i,k=1,2,\ldots$ (or semiintegers). In other words, the single qudit state $j=0,1/2,1,3/2,2,\ldots$ can be mapped onto the density operator of the system containing the subsystems like the state of two qubits. Using this mapping, the notion of the separability and the entanglement was extended in \cite{Markovich3} to the case of the single qudit $X$-state with $j=3/2$. Also, the concurrence and the negativity were introduced for such system. Later, the analog of the correlation function  for the single qudit and the notion of a quantum steering for the system without subsystems was obtained in \cite{Mar9}. Thus, it is necessary to detect the presence of the entanglement both in the systems with the subsystems and without them.
\par The aim of the paper is to study the dependence of quantum correlations associated with the entanglement phenomenon in composite and indivisible systems on the eigenvectors of the system state density matrices. It is known, that the entanglement is not related directly to eigenvalues of the density matrix \cite{Chuang}. Also the entanglement properties depend on the reference frame determined by the eigenvectors of observables (Hermitian matrices). We use the fact that the same density operator (the same numerical density matrix) with some spectrum can be interpreted either as the density matrix of the separable state or the entangled state. The density matrix is transformed by the unitary matrix if one changes the reference frame using two different systems of basis vectors. This unitary matrix can be constructed as the set of the normalized eigenvectors of the density matrix organized as a set of the unitary matrix  columns. Starting from the diagonal density matrix which determines the separable state and applying all the unitary matrices associated with all the possible eigenvectors to obtain all the density matrices with the same spectrum, we investigate all the possible transformations of the separable states in the entangled states and vise versa.
\par The paper is organized as follows. In Sec.~\ref{sec:15_1} we briefly remind the notion of the entanglement and the separability for the quantum systems. The Peres-Horodecki criterion, the negativity and the concurrence are discussed in details. In the next section the unitary transformation is applied to the pure state to find such transformations that bring the state to the separable one. In Sec.~\ref{sec:15_3} the general state case is considered. Several examples of special unitary rotation matrices are obtained. The domains of the rotation matrix elements corresponding to the transformations to the separable states are deduced. Finally,  in Sec.~\ref{sec:15_4} all the results are illustrated on the examples. The mixed state is firstly transformed to the Werner state with the known entanglement domain. Then, with the second unitary transformation, the Werner state with the parameter domain corresponding to the entangled state is transformed to the state, where this domain corresponds to the separable state.
\section{Entanglement}\label{sec:15_1}
\par The entanglement is an important phenomenon of the quantum system. Thus, it is crucial to understand the nature of the entanglement and methods of its detection.
 By definition, the pure state $|\Psi\rangle$ is  called separable if it can be written as a single
tensor product of states in subsystems $A$ and $B$
\begin{eqnarray*}|\Psi\rangle=|\Psi^{A}\rangle\otimes|\Psi^{B}\rangle   \end{eqnarray*}
otherwise it is called entangled and has the following form
\begin{eqnarray*}|\Psi\rangle=\sum\limits_{ij}c_{ij}|\Psi^{A}_{i}\rangle\otimes|\Psi^{B}_{j}\rangle ,  \end{eqnarray*}
where $c_{ij}$ are some complex coefficients.
The well known example of the pure and separable state is $|\uparrow\uparrow\rangle = |\uparrow\rangle \otimes |\uparrow\rangle = |00\rangle$. As the example of the pure and the entangled states the Bell states can be given, e.g. singlet $\frac{1}{\sqrt{2}}(|\uparrow\downarrow\rangle-|\downarrow\uparrow\rangle$.
\\A mixed state $\rho$ is called
separable if it can be written as
\begin{eqnarray*}\rho=\sum\limits_{i} p_i|A_i\rangle\langle A_i|\otimes|B_i\rangle \langle B_i|=\sum\limits_{i} p_i\rho_{i}^{A}\otimes\rho_{i}^{B},\quad p_i\in[0,1],\quad \sum\limits_{i}p_i=1,
   \end{eqnarray*}
   where $|A_i\rangle$ and $|B_i\rangle$ are the state-vectors corresponding to the subsystems $A$ and $B$, defined on the Hilbert spaces $\mathcal{H}_A$ and $\mathcal{H}_B$, respectively.
  The $\rho^{A}$ and  $\rho^{B}$ are density matrices of subsystems and the $p$ are probabilities.
If the mixed state cannot be represented in the latter form it is entangled. In contrast to the pure state, the form of the entangled mixed state cannot be explicitly formulated.
Thus, it is a challenging problem to decide if a given state is separable or entangled.
\\ In modern science, the entanglement is mostly studied for the bipartite systems. The reduced density matrix $\rho^{A}$ for the subsystem $A$ is
\begin{eqnarray*}\rho^{A}=Tr_B( \rho^{AB}),
   \end{eqnarray*}
   where $ \rho^{AB}$ is the density matrix of the bipartite system. Similarly, $\rho^{B}=Tr_A( \rho^{AB})$ represents the partial trace over the second subsystem $B$.
It is known, that the state is entangled if the reduced density matrix represents the mixed state and is separable if the reduced density matrix remains the pure
state. Unfortunately we can use the partial trace entanglement criteria only for the pure states.
\subsection{The Peres-Horodecki criterion}
\par There are various ways in which entanglement can be detected. One of them  is the Peres-Horodecki criterion.
Unlike the partial trace method, this criterion can be used for both pure and mixed states. The idea of the method in the following.
If we have the general state that acts on the Hilbert space $H=\mathcal{H}_A\otimes\mathcal{H}_B$
\begin{eqnarray*}\rho=\sum\limits_{ijkl} p_{kl}^{ij}|i\rangle\langle j|\otimes|k\rangle \langle l|,
   \end{eqnarray*}
   then its partial transpose  with respect to the second party is the following
   \begin{eqnarray*}{\rho^{ppt}}^{B}=\sum\limits_{ijkl} p_{kl}^{ij}|i\rangle\langle j|\otimes(|k\rangle \langle l|)^{T}=\sum\limits_{ijkl} p_{kl}^{ij}|i\rangle\langle j|\otimes|l\rangle \langle k|=\sum\limits_{ijkl} p_{lk}^{ij}|i\rangle\langle j|\otimes|k\rangle \langle l|.
   \end{eqnarray*}
   If there is no entanglement, then the eigenvalues $ \lambda^{ppt} $ of the transposed matrix $ \rho^{ppt} $ are positive \cite{Horodecki}.
 \par Let the quantum state in the four-dimensional Hilber space $\mathcal{H}$ be described by the density matrix
\begin{eqnarray}\rho={\left(
                                 \begin{array}{cccc}
                                   \rho_{11}& \rho_{12}& \rho_{13}& \rho_{14}\\
                                   \rho_{21}& \rho_{22}& \rho_{23}& \rho_{24}\\
                                   \rho_{31}& \rho_{32}& \rho_{33}& \rho_{34}\\
                                   \rho_{41}& \rho_{42}& \rho_{43}& \rho_{44}\\
                                 \end{array}
                               \right)}\, \label{33_1}
                               \end{eqnarray}
such that  $\rho=\rho^{\dagger}$, $Tr\rho=1$  and its eigenvalues $\lambda_i,i=1,2,3,4$ are nonnegative.
For the system of two qubits and qubit-qutrit system the PPT criterion is both necessary and sufficient condition for separability.
In higher dimensions the latter method is sufficient but not necessary condition for the entanglement. If the original quantum system is entangled, then the eigenvalues of the  density matrix $\rho^{ppt}$ can be negative.
For example, if the two-qubit system is described by the $ X $ - matrix
\begin{eqnarray}\label{15_22}
\rho^{X}&=&\left(
                     \begin{array}{cccc}
                       \rho_{11} & 0 & 0 & \rho_{14}\\
                       0 & \rho_{22}& \rho_{23} & 0 \\
                       0 & \rho_{32} & \rho_{33} & 0 \\
                       \rho_{41} & 0 & 0 & \rho_{44} \\
                     \end{array}
                   \right)=\left(
                     \begin{array}{cccc}
                       \rho_{11} & 0 & 0 & \rho_{14}\\
                       0 & \rho_{22}& \rho_{23} & 0 \\
                       0 & \rho_{23}^{\ast} & \rho_{33} & 0 \\
                      \rho_{14}^{\ast} & 0 & 0 & \rho_{44} \\
                     \end{array}
                   \right),
\end{eqnarray}
then the PPT matrix has the form
\begin{eqnarray}\label{15_16}
\rho^{Xppt}&=&\left(
                     \begin{array}{cccc}
                       \rho_{11} & 0 & 0 & \rho_{23}\\
                       0 & \rho_{22}& \rho_{14} & 0 \\
                       0 & \rho_{14}^{\ast} & \rho_{33} & 0 \\
                      \rho_{23}^{\ast} & 0 & 0 & \rho_{44} \\
                     \end{array}
                   \right),
\end{eqnarray}
where $\rho_{11},\rho_{22},\rho_{33},\rho_{44}$ are positive reals and $\rho_{23},\rho_{14}$ are complex quantities.
The eigenvalues of the latter matrix are the following
\begin{eqnarray}&&l_1= e_1 - e_3,\quad  l_2 =e_1 + e_3,\quad l_3=e_2 - e_4,\quad l_4=  e_2+ e_4,
\end{eqnarray}
where we introduced the  notations
\begin{eqnarray}\label{12_15}&&e_1=\frac{\rho_{11}+\rho_{44}}{2},\quad e_2=\frac{\rho_{22}+\rho_{33}}{2},\\\nonumber
&&e_3=\frac{\sqrt{\left(\rho_{11}-\rho_{44}\right)^2+4|\rho_{23}|^2}}{2},\quad e_4=\frac{\sqrt{\left(\rho_{22}-\rho_{33}\right)^2+4|\rho_{14}|^2}}{2}.\nonumber\end{eqnarray}
The latter matrix has the unit trace and it is nonnegative if $\rho_{22}\rho_{33}\geq|\rho_{23}|^2$, $\rho_{11}\rho_{44}\geq|\rho_{14}|^2$ hold.
Then, from the Peres-Horodecki criterion we can conclude that
\begin{eqnarray}\label{15_24}\frac{\rho_{11}+\rho_{44}}{2}\geq \frac{\sqrt{\left(\rho_{11}-\rho_{44}\right)^2+4|\rho_{23}|^2}}{2},
\frac{\rho_{22}+\rho_{33}}{2}\geq \frac{\sqrt{\left(\rho_{22}-\rho_{33}\right)^2+4|\rho_{14}|^2}}{2}.
\end{eqnarray}
\par  Note, that we did not specify what kind of the system describes the density matrix (\ref{33_1}). Diagonal elements of (\ref{33_1}) can be considered as components of the probability vector $\overrightarrow{p}=(p_{11},p_{22},p_{33},p_{44})$, $\sum\limits_{i}p_{ii}=1$, $0\leq p_{ii}\leq1$.
Let us use the invertible mapping of indices $1\leftrightarrow 1/2~1/2$; $2\leftrightarrow1/2~-1/2$; $3\leftrightarrow-1/2~1/2$; $4\leftrightarrow-1/2~-1/2$. For example, we formally recall the element $\rho_{11}$ as $\rho_{\frac{1}{2},\frac{1}{2}}$.  In the new notation (\ref{33_1}) can describe the two-qubit state.
On the other hand,  applying the invertible map of indices $1\leftrightarrow 3/2$, $2\leftrightarrow1/2$, $3\leftrightarrow-1/2$, $4\leftrightarrow-3/2$, the latter density matrix can be rewritten as
 \begin{eqnarray}\rho_{\frac{3}{2}}=\left(
                                 \begin{array}{cccc}
                                   \rho_{\frac{3}{2},\frac{3}{2}}& \rho_{\frac{3}{2},\frac{1}{2}}& \rho_{\frac{3}{2},-\frac{1}{2}}& \rho_{\frac{3}{2},-\frac{3}{2}}\\
                                   \rho_{\frac{1}{2},\frac{3}{2}}& \rho_{\frac{1}{2},\frac{1}{2}}& \rho_{\frac{1}{2},-\frac{1}{2}}& \rho_{\frac{1}{2},-\frac{3}{2}}\\
                                   \rho_{-\frac{1}{2},\frac{3}{2}}& \rho_{-\frac{1}{2},\frac{1}{2}}& \rho_{-\frac{1}{2},-\frac{1}{2}}& \rho_{-\frac{1}{2},-\frac{3}{2}}\\
                                   \rho_{-\frac{3}{2},\frac{3}{2}}& \rho_{-\frac{3}{2},\frac{1}{2}}& \rho_{-\frac{3}{2},-\frac{1}{2}}& \rho_{-\frac{3}{2},-\frac{3}{2}}\\
                                 \end{array}
                               \right)\label{33_6}
                               \end{eqnarray}
and can describe the noncomposite system of the single qudit with the spin $j=3/2$ \cite{Markovich3}.
The matrix keeps the standard properties of the density matrix, e.g., $\rho_{3/2}=\rho_{3/2}^{\dagger}$, $Tr\rho_{3/2}=1$ and its eigenvalues are nonnegative.
 It means that all  equalities and inequalities known for (\ref{33_1}) (the two-qubit system) are valid for (\ref{33_6}) (the single qudit system).
In future, we will not indicate what system the density matrix (\ref{33_1}) describes: the two-qubit or the single qudit one, but all the results obtained are valid for both quantum systems.
\subsection{Unitary matrix rotation}
\par Let us call $\rho_d$ the diagonal matrix with elements equal to the  eigenvalues of  (\ref{33_1}). Next, we
introduce the transformation $\rho_W = W\rho_dW^{\dagger}$, where  the unitary transform $W$ have the following matrix
\begin{eqnarray}\label{12_7}W = \left(
                       \begin{array}{cccc}
                         u_{11} & u_{12} & u_{13} & u_{14} \\
                         u_{21} & u_{22} & u_{23} & u_{24} \\
                         u_{31}& u_{32} & u_{33} & u_{34}\\
                         u_{41} & u_{42} & u_{43} & u_{44} \\
                       \end{array}
                     \right).
\end{eqnarray}
Since the matrix is unitary, the special conditions on its entries
\begin{eqnarray}\label{15_20}&&\sum_{i}|u_{ij}|^2=1,\quad\sum_{j}|u_{ij}|^2=1,\\\nonumber
 &&\prod\limits_{i}u_{i1}^{*}u_{ij}=0,\quad j=2,3,4,\quad  \prod\limits_{i}u_{i2}^{*}u_{ik}=0,\quad k=3,4\quad \prod\limits_{i}u_{i3}^{*}u_{i4}=0,\\\nonumber
  &&\prod\limits_{i}u_{1i}^{*}u_{ji}=0,\quad j=2,3,4,\quad  \prod\limits_{i}u_{2i}^{*}u_{ki}=0,\quad k=3,4\quad \prod\limits_{i}u_{3i}^{*}u_{4i}=0,\\nonumber
\end{eqnarray}
hold. For the matrix $\rho_W$ the  partial density matrices can be defined as
\begin{eqnarray*}\rho_W(1)=\left(
                  \begin{array}{cc}
                    \rho_{11}^W+\rho_{22}^W &\rho_{13}^W+\rho_{24}^W \\
                    \rho_{31}^W+\rho_{42}^W & \rho_{33}^W+\rho_{44}^W \\
                  \end{array}
                \right),\quad\rho_W(2)=\left(
                  \begin{array}{cc}
                    \rho_{11}^W+\rho_{33}^W &\rho_{12}^W+\rho_{34}^W \\
                    \rho_{21}^W+\rho_{43}^W & \rho_{22}^W+\rho_{44}^W \\
                  \end{array}
                \right)
                               \end{eqnarray*}
and we can write purity parameters for the latter subsystems as
 \begin{eqnarray*}\mu_1=1-Tr\rho_W(1)^2,\quad \mu_2=1-Tr\rho_W(2)^2.\end{eqnarray*}
 It is known, that for the pure state $\mu_1=\mu_2$.
\par Conditions (\ref{15_20}) are quite complicated, especially with the increasing of the dimensionality of the density matrix. Hence, sometimes it is convenient to use the parametrization of the unitary matrices introduced in \cite{Diji} that takes into account all the latter conditions, e.g., for the $4\times4$ unitary matrix the following parametrization for the elements of the first column of the rotation unitary matrix
\begin{eqnarray}\label{15_23}
u_{11}&=& a\exp(i\varphi_{11}),\\\nonumber
u_{21}&=& d\sqrt{1-a^2}\exp(i\varphi_{21}),\\\nonumber
u_{31}&=&f\sqrt{(1-a^2)(1-d^2)}\exp(i\varphi_{31}),\\\nonumber
u_{41}&=&\sqrt{(1-a^2)(1-d^2)(1-f^2)}\exp(i\varphi_{41})\nonumber
\end{eqnarray}
hold. The other matrix elements are given in Appendix. The latter parametrization can be introduced for the matrix of any dimension.  That can be very useful for the unitary transformation of the high-order density matrices. The matrix $W$ can transform the matrix $\rho_d$ to any matrix we like depending on the matrix entries. This unitary matrix can be constructed as the set of the normalized eigenvectors of the density matrix organized as a set of the unitary matrix columns. We will  start from the diagonal matrix $\rho_d$ constructed from the eigenvalues of the density matrix which determines the separable state. Then, applying all the unitary matrices associated with all the possible eigenvectors we will obtain all the density matrices with the same spectrum.
It is interesting to find such domains of the parameters $u_{ij}$ of the rotation matrix  (or the domains of the parameters (\ref{15_23})), where the Peres-Horodecki criterion breaks and the transformed state becomes entangled.
\subsection{Negativity and concurrence}
There exist other methods of the entanglement detection in the system called the concurrence and the negativity. Both of these methods are in some way connected with the eigenvalues of the density matrix of the state. Obviously, the eigenvalues of (\ref{15_22}) are
\begin{eqnarray*}\lambda_1^{ppt} &=&e_1+e_3,\quad\lambda_2^{ppt} =e_1-e_3,\quad
\lambda_3^{ppt} =e_2+e_4,\quad\lambda_4^{ppt}=e_2-e_4,
\end{eqnarray*}
where we use notations (\ref{12_15}).
In case where the following inequality
\begin{eqnarray}\label{15_39}&&|\lambda_1^{ppt}|+|\lambda_2^{ppt}|+|\lambda_3^{ppt}|+|\lambda_4^{ppt}|-1>0
\end{eqnarray}
 holds, the sum in the left hand side of this inequality is called the negativity parameter characterizing the quantum state with the density matrix $\rho$. The state with the density matrix satisfying  the latter inequality is entangled. It is known, that the $X$-state of two qubits is entangled if either $\rho_{22}\rho_{33}<|\rho_{14}|^2$ or $\rho_{11}\rho_{44}<|\rho_{23}|^2$ hold. Both conditions cannot be fulfilled simultaneously \cite{Mazhar}. For the single qudit state with the spin $j=3/2$ the similar conditions are introduced in \cite{Markovich3}.
\par The concurrence is defined as
 \begin{eqnarray}\label{C}C(\rho)= \max\{0,\sqrt{\lambda_1}-\sqrt{\lambda_2}-\sqrt{\lambda_3}-\sqrt{\lambda_4}\},
 \end{eqnarray}
 where $\lambda_i,i=1,2,3,4$ are the square-roots of the eigenvalues of matrix $\rho^{X}\widetilde{\rho}^{X}$ in decreasing order.
 If $C(\rho)> 0$, then the system displays the pairwise entanglement. The matrix $\widetilde{\rho}^{X}$ is obtained by spin flip operation on the density matrix (\ref{15_22})
                               \begin{eqnarray}\label{15_33}
\widetilde{\rho}_{3/2}^{X}&=&(\sigma_y\otimes\sigma_y)\rho^{X\ast}(\sigma_y\otimes\sigma_y),
\end{eqnarray}
where $\sigma_y$ is a Pauli matrix and
\begin{eqnarray*}(\sigma_y\otimes\sigma_y)&=&\left(
                     \begin{array}{cccc}
                       0 & 0 & 0 &-1\\
                       0 &0& 1 & 0 \\
                       0 & 1& 0 & 0 \\
                      -1 & 0 & 0 &0 \\
                     \end{array}
                   \right).
\end{eqnarray*}
Hence, (\ref{15_33}) is
\begin{eqnarray}\label{15_55}
\widetilde{\rho}_{3/2}^{X}&=&\left(
                     \begin{array}{cccc}
                       \rho_{44} & 0 & 0 & \rho_{14}\\
                       0 & \rho_{33}& \rho_{23} & 0 \\
                       0 & \rho_{23}^{\ast} & \rho_{22} & 0 \\
                      \rho_{14}^{\ast} & 0 & 0 & \rho_{11} \\
                     \end{array}
                   \right).
\end{eqnarray}
Multiplying matrices (\ref{15_22}) and (\ref{15_55}) we get the matrix
\begin{eqnarray*}
\rho_{3/2}^{X}\widetilde{\rho}_{3/2}^{X}&=&\left(
                     \begin{array}{cccc}
                       |\rho_{14}|^2+\rho_{11}\rho_{44} & 0 & 0 & 2\rho_{11}\rho_{14}\\
                       0 & |\rho_{23}|^2+\rho_{22}\rho_{33}& 2\rho_{22}\rho_{23} & 0 \\
                       0 & 2\rho_{33}\rho_{23}^{\ast} & |\rho_{23}|^2+\rho_{22}\rho_{33} & 0 \\
                      2\rho_{44}\rho_{14}^{\ast} & 0 & 0 & |\rho_{14}|^2+\rho_{11}\rho_{44} \\
                     \end{array}
                   \right),
\end{eqnarray*}
which has the following  eigenvalues
\begin{eqnarray*}\lambda_1^{C} &=&|\rho_{14}|^2+\rho_{11}\rho_{44}-2\sqrt{\rho_{11}\rho_{44}|\rho_{14}|^2}=(|\rho_{14}|-\sqrt{\rho_{11}\rho_{44}})^2,\\
\lambda_2^{C} &=&|\rho_{14}|^2+\rho_{11}\rho_{44}+2\sqrt{\rho_{11}\rho_{44}|\rho_{14}|^2}=(|\rho_{14}|+\sqrt{\rho_{11}\rho_{44}})^2,\\
\lambda_3^{C} &=&|\rho_{23}|^2+\rho_{22}\rho_{33}-2\sqrt{\rho_{22}\rho_{33}|\rho_{23}|^2}=(|\rho_{23}|-\sqrt{\rho_{22}\rho_{33}})^2,\\
\lambda_4^{C}&=&|\rho_{23}|^2+\rho_{22}\rho_{33}+2\sqrt{\rho_{22}\rho_{33}|\rho_{23}|^2}=(|\rho_{23}|+\sqrt{\rho_{22}\rho_{33}})^2.
\end{eqnarray*}
Hence, it can be deduced that the concurrence (\ref{C}) is determined by
 \begin{eqnarray*}C(\rho)= \max\{0,2|\rho_{23}|-2\sqrt{\rho_{11}\rho_{44}},2|\rho_{14}|-2\sqrt{\rho_{22}\rho_{33}}\},
 \end{eqnarray*}
see for example \cite{Hedemann}. Note that the concurrence works only for $2\times 2$ dimensional systems, while the negativity being based on the PPT criterion is not able to capture all entanglement for systems of dimension greater than $2 \times 3$.
The negativity and the concurrence can be written for $\rho_W$ and provide different conditions on the rotation matrix $W$.

\section{The pure state case}\label{sec:15_2}
\par First, we investigate the case when $\rho_d$ is the diagonal matrix with $\rho_{11}=1$ and other matrix elements are zero, namely, the pure state.
 Using (\ref{12_7}) we can write the transformed matrix as
 \begin{eqnarray}\label{12_18}\rho_W =W\rho_dW^{\dagger}= \left(
                       \begin{array}{cccc}
                         |u_{11}|^2 &u_{21}u_{11}^{*} &u_{11}u_{31}^{*} & u_{21}u_{31}^{*}\\
                         u_{11}u_{21}^{*} & |u_{21}|^2  &  u_{11}u_{41}^{*} & u_{21}u_{41}^{*} \\
                        u_{31}u_{11}^{*}& u_{41}u_{11}^{*} & |u_{31}|^2 &u_{41}u_{31}^{*}\\
                        u_{31}u_{21}^{*}  &u_{41}u_{21}^{*}  &  u_{31}u_{41}^{*}& |u_{41}|^2  \\
                       \end{array},
                     \right)
\end{eqnarray}
depending only on the first column of the rotation matrix.
Next, after the PPT of the latter matrix, we can find its eigenvalues, namely,
 \begin{eqnarray}\label{12_19}\lambda_{W1,2}^{ppt}&=& \pm u_{11}^{*}u_{41}^{*} \mp u_{21}^{*}u_{31}^{*}\sqrt{u_{11}u_{41} - u_{21}u_{31}},\\\nonumber
 \lambda_{W3,4}^{ppt}&=&\frac{1}{2}\mp\frac{1}{2} \sqrt{1- 4\left(u_{11}u_{41} - u_{21}u_{31}\right) \left(u_{11}^{*}u_{41}^{*} - u_{21}^{*}u_{31}^{*}\right)}.
 \end{eqnarray}
The latter eigenvalues are nonnegative if
  \begin{eqnarray}\label{12_21}&&u_{11}^{*}u_{41}^{*}=u_{21}^{*}u_{31}^{*}\sqrt{u_{11}u_{41} - u_{21}u_{31}},\\\nonumber
&&1\geq\sqrt{1- 4\left(u_{11}u_{41} - u_{21}u_{31}\right) \left(u_{11}^{*}u_{41}^{*} - u_{21}^{*}u_{31}^{*}\right)}
 \end{eqnarray}
 hold.  Using the parametrization (\ref{15_23}) of the unitary matrix, we can find the domain of parameters of the unitary transformation matrix, where it does the transition to the separable state.  Substituting (\ref{15_23}) into (\ref{12_21}), we can deduce that
   \begin{eqnarray}\label{15_40}&&\sqrt{(1-d^2)(1-a^2)}\Bigg(ae^{-i(\varphi_{11}+\varphi_{41})}\sqrt{1-f^2}-df(1-a^2)\sqrt{1-d^2}e^{-i(\varphi_{21}+\varphi_{31})}\nonumber\\
   &\times&   \sqrt{-df\sqrt{1-a^2}e^{i(\varphi_{21}+\varphi_{31})}+a\sqrt{(1-f^2)}e^{i(\varphi_{11}+\varphi_{41})}}\Bigg)=0.
   \end{eqnarray}
Finally, from the latter condition we can conclude that
   \begin{eqnarray}
   &&1\geq\Bigg(1-4(1-d^2)(1-a^2)(a^2(1-f^2)+d^2f^2(1-a^2)\\\nonumber
   &-&2adf\sqrt{(1-f^2)(1-a^2)}\cos(\varphi_{11}-\varphi_{21}-\varphi_{31}+\varphi_{41})\Bigg)^{1/2}.
 \end{eqnarray}
The latter inequality holds if
\begin{eqnarray}\label{15_12}&&a=1,\quad \forall d,f, \varphi_{ij};\quad d=1,\quad \forall a,f, \varphi_{ij};\\\nonumber
&&a=0,\quad d=0\quad\mbox{or}\quad f=0,\quad\forall \varphi_{ij};\\\nonumber
&&a=0,\quad d=\frac{a\sqrt{1-f^2}e^{i(\varphi_{11}+\varphi_{41})}}{f\sqrt{1-a^2}e^{i(\varphi_{21}+\varphi_{31})}},\quad\forall \varphi_{ij};\\\nonumber
&&f=1,\quad d=0,\quad \forall a,\varphi_{ij}.
\end{eqnarray}
For all the  other parameters the rotation matrix $W$ transfers the state to the entangled one.
\section{General state case}\label{sec:15_3}
\par Let  $\rho_d$ be the diagonal matrix with the elements $\rho_{ii}=l_i$, $i=1,2,3,4$, where $l_i$  are the eigenvalues of the density matrix (\ref{33_1}).
After the $W$ transformation and the PPT, the matrix is the following
 \begin{eqnarray}\label{12_23}\rho_W^{ppt} = \left(
                       \begin{array}{cccc}
                         \rho_{11} &\rho_{12}&\rho_{13} & \rho_{14}\\
                         \rho_{12}^{*} &  \rho_{22} &   \rho_{23}&  \rho_{24}\\
                         \rho_{13}^{*}&  \rho_{23}^{*} & \rho_{33} & \rho_{34}\\
                         \rho_{14}^{*}  & \rho_{24}^{*} &  \rho_{34}^{*}&  \rho_{44} \\
                       \end{array}
                     \right),
\end{eqnarray}
where $\rho_{ij}$ are given in Appendix.
The latter matrix is Hermitian and the equation on  the eigenvalues of the PPT matrix $\rho_W^{ppt}$ is
\begin{eqnarray}\label{15_6}\lambda^4+a_1\lambda^3+a_2\lambda^2+a_3\lambda+a_4=0,
\end{eqnarray}
where the coefficients $a_i$ depend on (\ref{15_25_1}).
Since the eigenvalues of the latter matrix are rather big we shall illustrate our idea on examples.
\\Let us select the matrix (\ref{12_7}) of the following type
\begin{eqnarray}\label{12_14}W = \left(
                       \begin{array}{cccc}
                         u_{11} & 0 & u_{13} & 0 \\
                         0 & u_{22} & 0 & u_{24} \\
                         u_{31}& 0 & u_{33} & 0\\
                         0 & u_{42} & 0 & u_{44} \\
                       \end{array}
                     \right),\quad WW^{\dagger}=I.
\end{eqnarray}
Hence, from the unitary conditions we can write
\begin{eqnarray}\label{12_16}&&|u_{11}|^2+|u_{13}|^2=1,\quad |u_{31}|^2+|u_{33}|^2=1,\quad |u_{13}|^2+|u_{33}|^2=1\\ \nonumber
&&|u_{22}|^2+|u_{24}|^2=1,\quad |u_{42}|^2+|u_{44}|^2=1,\quad|u_{24}|^2+|u_{44}|^2=1\\ \nonumber
&&|u_{11}|^2+|u_{31}|^2=1,\quad|u_{22}|^2+|u_{42}|^2=1,\quad u_{22}u_{42}^{*}+u_{24}u_{44}^{*}=0\\ \nonumber
&&u_{11}u_{13}^{*}+u_{31}u_{33}^{*}=0,\quad u_{11}u_{31}^{*}+u_{13}u_{33}^{*}=0,\quad u_{22}u_{24}^{*}+u_{42}u_{44}^{*}=0 .
\nonumber
\end{eqnarray}
Thus, it is easy to verify that
\begin{eqnarray}\label{12_26}&&
|u_{13}|^2=|u_{31}|^2,\quad |u_{24}|^2=|u_{42}|^2,\quad |u_{22}|^2=|u_{44}|^2,\quad |u_{11}|^2=|u_{33}|^2.
\end{eqnarray}
The latter matrix transforms $\rho_d$ into the following one
\begin{eqnarray}\label{12_24}
\rho_{W}=W\rho_dW^{\dagger}=\left(
                                  \begin{array}{cccc}
                                    \rho_{11} &0 & \rho_{13} & 0 \\
                                    0&  \rho_{22} &0& \rho_{24} \\
                                    \rho_{31} & 0 & \rho_{33}& 0 \\
                                    0 & \rho_{42} & 0& \rho_{44} \\
                                  \end{array}
                                \right),
\end{eqnarray}
where $\rho_{ij}$ are given in Appendix.
If the latter matrix is the density matrix, then it always corresponds to the separable state. From the definition of the density matrix
one can deduce the following constrains on its  elements
\begin{eqnarray*}&&\sum\limits_{i}\rho_{ii}=1,\quad \rho_{13}=\rho_{31}^{*},\quad \rho_{24}=\rho_{42}^{*},\\
&& \frac{\rho_{11}+\rho_{33}}{2}\pm\frac{\sqrt{\left(\rho_{11}-\rho_{33}\right)^2+4|\rho_{13}|^2}}{2}\geq0,\\
&& \frac{\rho_{22}+\rho_{44}}{2}\pm\frac{\sqrt{\left(\rho_{22}-\rho_{44}\right)^2+4|\rho_{24}|^2}}{2}\geq0.
\end{eqnarray*}
Hence, we can write the following conditions
\begin{eqnarray*}l_1(|u_{11}|^2+|u_{13}|^2)+l_2(|u_{22}|^2+|u_{24}|^2)+l_3(|u_{13}|^2+|u_{33}|^2)+l_4(|u_{44}|^2+|u_{24}|^2)=1
\end{eqnarray*}
and
\begin{eqnarray*}
&& \frac{l_1(|u_{11}|^2+ |u_{31}|^2)+l_3(|u_{13}|^2+|u_{33}|^2)}{2}\pm\\
&\pm&\frac{\sqrt{\left(l_1(|u_{11}|^2- |u_{31}|^2)+l_3(|u_{13}|^2-|u_{33}|^2)\right)^2+4|l_1u_{11}u_{31}^{*}+l_3u_{13}u_{33}^{*}|^2}}{2}\geq0,\\
&& \frac{l_4(|u_{24}|^2+|u_{44}|^2)+l_2(|u_{22}|^2+|u_{42}|^2)}{2}\pm\\
&\pm&\frac{\sqrt{\left(l_4(|u_{24}|^2- |u_{44}|^2)+l_2(|u_{22}|^2-|u_{42}|^2)\right)^2+4|l_2u_{22}u_{42}^{*}+ l_4u_{24}u_{44}^{*}|^2}}{2}\geq0.\\
\end{eqnarray*}
Let us sum up the two latter inequalities and, taking into account the condition (\ref{12_16}),  we can write the following extra condition to conditions (\ref{12_16}) on the elements of the unitary matrix $W$ defined by (\ref{12_14})
\begin{eqnarray*}
&&\sqrt{l_1^2+l_3^2+2l_1l_3\left(-(2|u_{11}|^2-1)^2+4u_{11}u_{33}(1-|u_{33}|^2)\right)}\\
&+&\sqrt{l_2^2+l_4^2+2l_2l_4\left(-(2|u_{22}|^2-1)^2+4u_{22}u_{44}(1-|u_{44}|^2)\right)}\leq1.\\
\end{eqnarray*}
\\The next example is the unitary matrix $W$ in the block form
\begin{eqnarray}\label{12_14}W = \left(
                       \begin{array}{cccc}
                         u_{11} & u_{12} & 0 & 0 \\
                        u_{21}& u_{22} & 0 & 0 \\
                         0& 0 & u_{33} & u_{34}\\
                         0 & 0 & u_{43} & u_{44} \\
                       \end{array}
                     \right),\quad WW^{\dagger}=I.
\end{eqnarray}
From the unitary condition
\begin{eqnarray*}&|u_{11}|^2+|u_{21}|^2=1,\quad |u_{22}|^2+|u_{21}|^2=1,\quad |u_{34}|^2+|u_{33}|^2=1,\quad |u_{43}|^2+|u_{44}|^2=1,\\
&|u_{11}|^2+|u_{12}|^2=1,\quad |u_{22}|^2+|u_{12}|^2=1,\quad |u_{43}|^2+|u_{33}|^2=1,\quad |u_{34}|^2+|u_{44}|^2=1\nonumber
\end{eqnarray*}
hold. Hence, we can deduce the following constrains on the elements of the unitary matrix $W$
\begin{eqnarray}\label{12_27}&&
|u_{21}|^2=|u_{12}|^2,\quad |u_{34}|^2=|u_{43}|^2,\quad |u_{11}|^2=|u_{22}|^2,\quad |u_{33}|^2=|u_{44}|^2.
\end{eqnarray}
The following matrix can be written
\begin{eqnarray}\label{12_25}
\rho_{W}=W\rho_dW^{\dagger}=\left(
                                  \begin{array}{cccc}
                                    \rho_{11} & \rho_{12} & 0 & 0\\
                                     \rho_{21}&  \rho_{22} &0& 0\\
                                    0 & 0 & \rho_{33}& \rho_{34} \\
                                    0 & 0 & \rho_{43}& \rho_{44} \\
                                  \end{array}
                                \right),
\end{eqnarray}
where $\rho_{ij}$ are given in Appendix.
Similar to the previous case, we can conclude the following condition on the matrix elements of the unitary rotation matrix $W$
\begin{eqnarray*}&&\sqrt{((2|u_{11}|^2-1)(l_1-l_2))^2
+4(1-|u_{11}|^2)|l_1u_{11}+l_2u_{22}|^2}\\
&+&\sqrt{((2|u_{33}|^2-1)(l_3- l_4))^2
+4(1-|u_{33}|^2)|l_3u_{33}+  l_4u_{44}|^2}\leq1.\nonumber
\end{eqnarray*}
\\ Obviously, there exists the unitary matrix $W$ such that
\begin{eqnarray*}\rho^{X}&=&W\rho_dW^{\dagger}.
\end{eqnarray*}
Let us select the matrix (\ref{12_7}) of the following $X$-type
\begin{eqnarray}\label{15_36}W = \left(
                       \begin{array}{cccc}
                         u_{11} & 0 & 0 & u_{14} \\
                        0 & u_{22} & u_{23} & 0 \\
                         0& u_{32} & u_{33} & 0\\
                         u_{41} & 0 & 0 & u_{44} \\
                       \end{array}.
                     \right),\quad WW^{\dagger}=I.
\end{eqnarray}
Since the matrix is unitary, we can conclude that
\begin{eqnarray}\label{12_17}&|u_{11}|^2+|u_{41}|^2=1,\quad |u_{22}|^2+|u_{32}|^2=1,\quad |u_{23}|^2+|u_{33}|^2=1,\\\nonumber
& |u_{14}|^2+|u_{44}|^2=1,\quad |u_{11}|^2+|u_{14}|^2=1,\quad |u_{22}|^2+|u_{23}|^2=1,\\\nonumber
&|u_{32}|^2+|u_{33}|^2=1,\quad |u_{41}|^2+|u_{44}|^2=1,\nonumber
\end{eqnarray}
\begin{eqnarray}\label{12_29}
&&u_{11}u_{14}^{*}+u_{41}u_{44}^{*}=0,\quad u_{11}u_{41}^{*}+u_{14}u_{44}^{*}=0,\\\nonumber
&& u_{22}u_{32}^{*}+u_{23}u_{33}^{*}=0,\quad u_{22}u_{23}^{*}+u_{32}u_{33}^{*}=0.
\end{eqnarray}
and we can conclude the following constrains
\begin{eqnarray}\label{12_28}&&
|u_{23}|^2=|u_{32}|^2,\quad |u_{14}|^2=|u_{41}|^2,\quad |u_{11}|^2=|u_{44}|^2,\quad |u_{22}|^2=|u_{33}|^2.
\end{eqnarray}
The latter matrix transforms $\rho_d$ into the $X$-matrix
\begin{eqnarray}\label{15_30}
\rho_{X_W}=W\rho_dW^{\dagger}=\left(
                                  \begin{array}{cccc}
                                    \rho_{11} &0 & 0 & \rho_{14} \\
                                    0&  \rho_{22} &\rho_{23}& 0\\
                                    0 & \rho_{32} & \rho_{33}& 0 \\
                                    \rho_{41} & 0 & 0& \rho_{44} \\
                                  \end{array}
                                \right)
\end{eqnarray}
where $\rho_{ij}$ are given in Appendix.
Similarly to the previous examples, we can write the following inequality
\begin{eqnarray}\label{15_34}
&&\sqrt{\left((2|u_{11}|^2-1)(l_1-l_4)\right)^2+4(1-|u_{22}|^2)(l_2u_{22}+l_3u_{33})^2}\\\nonumber
&+&\sqrt{\left((2|u_{22}|^2-1)(l_2-l_3)\right)^2+4(1-|u_{11}|^2)(l_1u_{11}+l_4u_{44})^2}\leq1.
\end{eqnarray}
From the Peres-Horodecki criterium the latter state is separable if (\ref{15_24}) holds, e.g.,
\begin{eqnarray}&&\label{15_26}\frac{l_1+l_4}{2}\geq \frac{\sqrt{\left((l_1-l_4)(2|u_{11}|^2-1)\right)^2+4|l_2u_{22}u_{32}^{*}+ l_3u_{23}u_{33}^{*}|^2}}{2},\\\nonumber
&&\frac{l_3+l_2}{2}\geq \frac{\sqrt{\left((l_2-l_3)(2|u_{22}|^2-1)\right)^2+4|l_1u_{11}u_{41}^{*}+l_4u_{14}u_{44}^{*}|^2}}{2}.
\end{eqnarray}
The latter inequalities provide an extra condition to (\ref{12_29}) on the rotation unitary matrix $W$ defined by (\ref{12_14}) under which it transforms the matrix $\rho_d$ to the separable $X$-state.
\section{Examples}\label{sec:15_4}
\begin{figure}
\begin{center}
\includegraphics[width=0.75\textwidth]{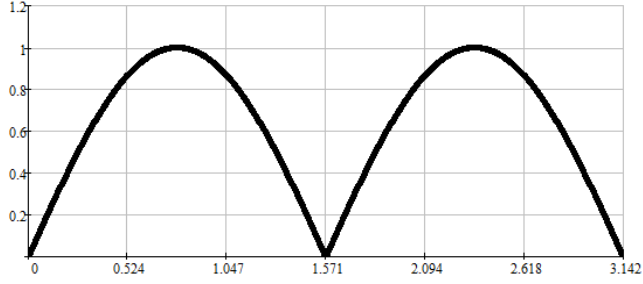}
\caption{The negativity for the matrix (\ref{15_41}) against $\varphi\in[0,\pi]$.}\label{fig:15_2}
\end{center}
\end{figure}

\begin{figure}
\begin{center}
\includegraphics[width=0.75\textwidth]{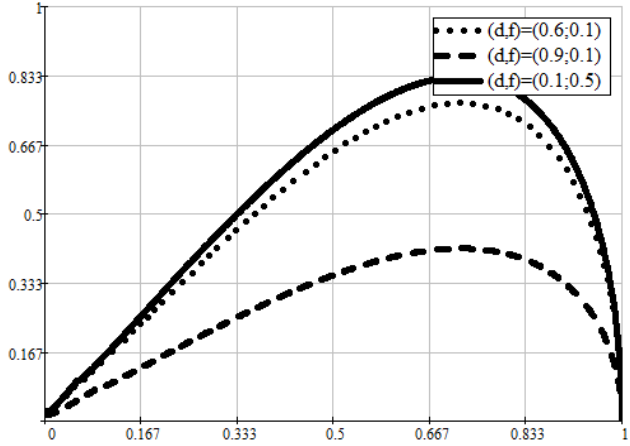}
\caption{The negativity for the matrix (\ref{12_18}) against $a\in[0,1]$.}\label{fig:15_3}
\end{center}
\end{figure}

In this section we illustrate the results obtained by several examples. First, we select the pure state described in Sec.~\ref{sec:15_2} by the matrix  $\rho_d$
with the diagonal elements $\{1,0,0,0\}$. The rotation matrix (\ref{15_30}) of the special view
\begin{eqnarray*}W = \left(
                       \begin{array}{cccc}
                         u_{11} & 0 & 0 & u_{14} \\
                        0 & 1 & 0 & 0 \\
                         0& 0 &1 & 0\\
                         u_{41} & 0 & 0 & u_{44} \\
                       \end{array}
                     \right),\quad WW^{\dagger}=I
\end{eqnarray*}
 transforms $\rho_d$ into the following matrix
\begin{eqnarray*}\rho_{W}= \left(
                       \begin{array}{cccc}
                         |u_{11}|^2 & 0 & 0 & u_{11}u_{41}^{*} \\
                        0 & 1 & 0 & 0 \\
                         0& 0 & 1 & 0\\
                         u_{41}u_{11}^{*}  & 0 & 0 & |u_{41}|^2 \\
                       \end{array}
                     \right).
\end{eqnarray*}
After the PPT of the latter matrix it has four eigenvalues
\begin{eqnarray}\label{15_38}&&\lambda_1=|u_{11}|^2,\quad \lambda_2=|u_{41}|^2,\quad\lambda_3=|u_{41}||u_{11}|,\quad\lambda_4=-|u_{41}||u_{11}|.
\end{eqnarray}
From the positivity condition the state is separable if $u_{11}=0$ or $u_{41}=0$ holds. For the other matrix elements of $W$ the pure state is transformed into the entangled state.
In the special case, when the rotation matrix depends only on one parameter $\varphi$, e.g.,
\begin{eqnarray}\label{15_41}W(\varphi)= \left(
                       \begin{array}{cccc}
                         \cos{\varphi} & 0 & 0 & \sin{\varphi} \\
                        0 & 1& 0 & 0 \\
                         0& 0 &1 & 0\\
                        -\sin{\varphi}  & 0 & 0 & \cos{\varphi}\\
                       \end{array}
                     \right)
\end{eqnarray}
the eigenvalues (\ref{15_38}) can be written explicitly as
\begin{eqnarray*}&&\lambda_1= \cos{\varphi}^2,\quad \lambda_2=\sin{\varphi} ^2,\quad\lambda_3=\sin{\varphi} \cos{\varphi} ,\quad\lambda_4=-\sin{\varphi} \cos{\varphi}.
\end{eqnarray*}
Hence, the state is separable if $\varphi=\pi,\pi/2$, i.e.
\begin{eqnarray*}\rho_{W}(0)= \left(
                       \begin{array}{cccc}
                         1& 0 & 0 &0 \\
                        0 & 0& 0 & 0 \\
                         0& 0 &0 & 0\\
                        0  & 0 & 0 & 0\\
                       \end{array}
                     \right),\quad \!\!\!\!\rho_{W}\left(\frac{\pi}{4}\right)= \frac{1}{2}\left(
                       \begin{array}{cccc}
                         1& 0 & 0 &-1 \\
                        0 & 0& 0 & 0 \\
                         0& 0 &0 & 0\\
                        -1  & 0 & 0 & 1\\
                       \end{array}
                     \right),\quad \!\!\!\!\rho_{W}(\pi)= \left(
                       \begin{array}{cccc}
                         0& 0 & 0 &0 \\
                        0 & 0& 0 & 0 \\
                         0& 0 &0 & 0\\
                        0  & 0 & 0 & 1\\
                       \end{array}
                     \right).
\end{eqnarray*}
It is visible that the first and the third matrix correspond to the separable states, but the second matrix not.
The negativity parameter (\ref{15_39}) depending on the rotation angle $\varphi$ is shown in Fig.~\ref{fig:15_2}.
It is visible, that the negativity is positive for all angles except of $\varphi=0,\pi,\pi/2$. That means that the matrix (\ref{15_41}) transforms the pure state $\rho_d$ in the separable state for only two angles $\varphi=\pi,\pi/2$. All the other rotations transfer the state in the entangled ones.
\\ More generally, for (\ref{12_18}) the negativity  against $a\in[0,1]$ is shown in Fig.~\ref{fig:15_3} for different values of parameters $(d,f)=\{(0.6,0.1);(0.9,0.1);(0.1,0.5)\}$ in the case when $\varphi_{i1}=0,i=1,2,3,4$.
As it is shown in (\ref{15_12}), in this case the matrix $W$ transforms the pure state $\rho_d$ in the separable one only for $a=0$ and $a=1$.
\par
Now, let us turn to the mixed state $\rho_d$ described in Sec.~\ref{sec:15_3} with the diagonal elements $l_i$, $i=1,2,3,4$. As the rotation matrix we select the $X$-type (\ref{15_36}) with
\begin{eqnarray}\label{15_43}&&u_{11}=\frac{p\pm\sqrt{p^2+8l_1(p+1)}}{2(p+1)},\quad u_{22}=\pm\frac{2\sqrt{l_2}}{\sqrt{1-p}},\\\nonumber
&&u_{33}=\pm\frac{2\sqrt{l_3}}{\sqrt{1-p}},\quad u_{44}=\frac{-p\pm\sqrt{p^2+8l_4(p+1)}}{2(p+1)},\\\nonumber
&&u_{41}=-u{11},\quad u_{14}=u{44}.\nonumber
\end{eqnarray}
The other matrix elements are equal to zero. The latter matrix elements have to satisfy (\ref{12_28}) and (\ref{15_34}). Hence, the elements are $l_1=l_4=(p+1\pm\sqrt{2}p)/4$, $l_2=l_3=(1-p)/4$ and the rotation matrix with elements (\ref{15_43}) transforms the mixed state $\rho_d$ to the following one
\begin{eqnarray}\label{10_1}\rho^{Wer}&=&\left(
                     \begin{array}{cccc}
                       \frac{1+p}{4} & 0 & 0 & \frac{p}{2}\\
                       0 & \frac{1-p}{4}& 0 & 0 \\
                       0 & 0& \frac{1-p}{4} & 0 \\
                       \frac{p}{2} & 0 & 0 & \frac{1+p}{4} \\
                     \end{array}
                   \right).
\end{eqnarray}
The latter matrix describes  the Werner state \cite{Werner} and the parameter $p$ satisfies the inequality $-\frac{1}{3}\leq p\leq1$. It is known, that the parameter domain $\frac{1}{3}< p\leq1$ corresponds to the entangled state. Hence, we have shown that the rotation matrix with (\ref{15_43}) can transform the mixed state $\rho_d$ to the state that can be either entangled one or the separable one depending on the parameter domain.
\par Next, let us take the rotation matrix (\ref{15_41}). After the transformation of the Werner state (\ref{10_1}) the PPT matrix is
\begin{eqnarray*}\rho^{ppt}= \frac{1}{2}\left(
                       \begin{array}{cccc}
                        p+1+p\cos{2\varphi} & 0 & 0 & 0 \\
                        0 & 1-p& p\cos{2\varphi}& 0 \\
                         0& p\cos{2\varphi} &1-p & 0\\
                        0  & 0 & 0 &   p+1-p\cos{2\varphi}  \\
                       \end{array}
                     \right).
\end{eqnarray*}
\begin{figure}
\begin{center}
\includegraphics[width=0.75\textwidth]{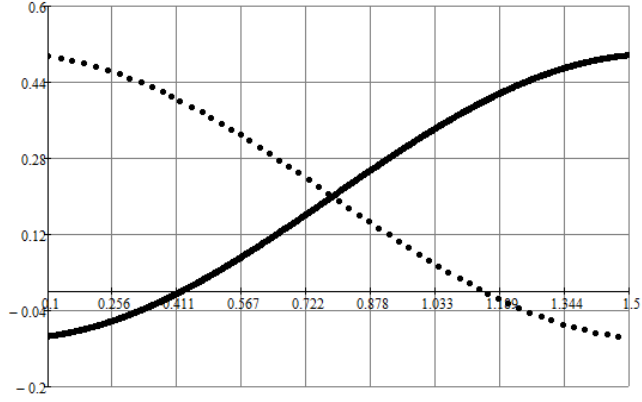}
\caption{The first (dotted line) and the second (black line) eigenvalues (\ref{15_44}) against $\varphi$.}\label{fig:15_4}
\end{center}
\end{figure}
Its eigenvalues are the following
\begin{eqnarray}\label{15_44}&&\lambda_{1,2}(p,\varphi)=\frac{1}{2}\left(1-p\pm p\cos{2\varphi} \right),\\
&&\lambda_{3}(p,\varphi)=\frac{1}{2}\left(1+p\pm p\sin{2\varphi} \right).\nonumber
\end{eqnarray}
From there positivity condition the following condition on the rotation matrix elements
\begin{eqnarray}\frac{\pi}{2}-\frac{1}{2}\arccos\left(\frac{p-1}{p}\right)<\varphi<\frac{1}{2}\arccos\left(\frac{p-1}{p}\right)\label{15_45}
\end{eqnarray}
hold. Hence, the eigenvalues are nonnegative in the parameter domain (\ref{15_45}). Since the argument of $\arccos(x)$ must be $-1\leq x \leq1$ it can be deduced that $p\in[1/2,1]$. The first and the second eigenvalues (\ref{15_44}) against $\varphi$  for $p=0.6$ are shown  in Fig.~\ref{fig:15_4}.
We select $p=0.6$ from the parameter domain, where the Werner state is entangled. The matrix $W$ can translate this entangled state into the separable one. In Fig.~\ref{fig:15_4}
this domain is $0.421<\varphi<1.15$.
\\ Thereby, taking a mixed state $\rho_d$, we first transfer it to the  Werner state, where the parameter domains corresponding to the entanglement and the separability are known. Next, using the second unitary transformation, we transfer the entangled Werner state in the new separable state.
\section{Conclusion}
Let us summarize the results obtained.
The results of the paper demonstrate the connection  of the quantum correlations associated with the entanglement phenomenon in the composite and the indivisible systems with the eigenvectors of the system state density matrices. Using the known fact, that  the entanglement properties depend on the reference frame determined by the eigenvectors of observables, we introduced the unitary matrix constructed  as the set of the normalized eigenvectors of the density matrix organized as a set of the unitary matrix  columns. Using the special parametrization of the unitary matrices introduced in \cite{Diji} we find the domain of parameters of the transformation matrix that transforms the pure states density matrix into the separable state. For the general case of the mixed states the latter domains were found for the special types of the rotation matrices like the cellular-matrix, the block-matrix and the $X$-matrix.
 All the results are illustrated on the examples of the special pure and mixed states and the rotation matrices of the different types. It would be interesting to extend the presented results to continuous variable (Gaussian) systems \cite{Simon} and going far afield to hybrid systems (continuous variable and discrete systems) \cite{ManciniMankoMankoTombesi}.

\begin{acknowledgements}
The study in section 3 and 4 by Markovich L.A. was supported by the Russian Science Foundation grant (14-50-00150).
\end{acknowledgements}

\bibliographystyle{spphys}       
\bibliography{fer_15}   

\section{Appendix}
\subsection{Appendix 1}
The parametrization of the unitary matrices (\ref{12_7}) introduced in \cite{Diji}
\begin{eqnarray*}u_{12}&=&b\sqrt{1-a^2}\exp(i\varphi_{22}),\\
u_{13}&=& c\sqrt{(1-a^2)(1-b^2)}\exp(i\varphi_{13}),\quad
u_{14}= \sqrt{(1-a^2)(1-b^2)(1-c^2)}\exp(i\varphi_{14}),\end{eqnarray*}
\begin{eqnarray*}
u_{22}&=&-abd\exp(i(\varphi_{12}+\varphi_{21}-\varphi_{11}))+\alpha\beta\sqrt{(1-b^2)(1-d^2)}\\
&\cdot&\Big(\sqrt{cfh\exp(i\varphi_{22})+cd(1-f^2)(1-h^2)}\exp(i\varphi_{32})\\
&+&b\sqrt{1-c^2}\exp(i\varphi_{23})(f\sqrt{1-h^2}-dh\sqrt{1-f^2}\exp(i(\varphi_{32}-\varphi_{22})))\Big),\end{eqnarray*}
\begin{eqnarray*}u_{23}&=&-acd\sqrt{1-b^2}\exp(i(\varphi_{21}+\varphi_{13}-\varphi_{11}))\\
&-&\alpha\beta\sqrt{1-d^2}\exp(i(\varphi_{13}-\varphi_{12}))(b(fh\exp(i\varphi_{22})\\
&+&d\sqrt{(1-f^2)(1-h^2)}\exp(i\varphi_{32}))-c(1-b^2)\sqrt{1-c^2}\exp(i\varphi_{23})\\
&\cdot&(f\sqrt{1-h^2}-dh\sqrt{1-f^2}\exp(i(\varphi_{32}-\varphi_{22})))),\end{eqnarray*}
\begin{eqnarray*}u_{24}&=&-ad\sqrt{(1-b^2)(1-c^2)}\exp(i(\varphi_{21}+\varphi_{41}-\varphi_{11}))\\
&-&\frac{\alpha}{\beta}\sqrt(1-d^2)\exp(i(\varphi_{14}+\varphi_{23}-\varphi_{12}))\\
&\cdot&(f\sqrt{1-h^2}-dh(1-f^2)\exp(i(\varphi_{32}-\varphi_{22}))),\end{eqnarray*}
\begin{eqnarray*}u_{32}&=&-abf\sqrt{1-d^2}\exp(i(\varphi_{12}+\varphi_{31}-\varphi_{11}))\\
&+&\alpha\beta\sqrt{1-b^2}\exp(-i\varphi_{21})(c(-dh\exp(i(\varphi_{22}+\varphi_{31}))\\
&+&f(1-d^2)\sqrt{(1-f^2)(1-h^2)}\exp(i(\varphi_{31}+\varphi_{32})))\\
&-&b\sqrt{1-c^2}\exp(i(\varphi_{23}+\varphi_{31}))(d\sqrt{1-h^2}+hf(1-d^2)\\
&\cdot&\sqrt{1-f^2}\exp(i(\varphi_{32}-\varphi_{22})))),\end{eqnarray*}
\begin{eqnarray*}u_{33}&=&-acf\sqrt{(1-b^2)(1-d^2)}\exp(i(\varphi_{13}+\varphi_{31}-\varphi_{11}))\\
&-&\alpha\beta\exp(i(\varphi_{13}-\varphi_{12}-\varphi_{21}))(-bdh\exp(i(\varphi_{31}+\varphi_{22}))\\
&+&bf(1-d^2)\sqrt{(1-f^2)(1-h^2)}\exp(i(\varphi_{31}+\varphi_{32}))\\
&+&c\sqrt{(1-b^2)(1-c^2)}\exp(i(\varphi_{23}+\varphi_{31}))(d\sqrt{1-h^2}\\
&+&hf(1-d^2)\sqrt{1-f^2}\exp(i(\varphi_{32}-\varphi_{22})))),\end{eqnarray*}
\begin{eqnarray*}u_{34}&=&-af\sqrt{(1-b^2)(1-c^2)(1-d^2)}\exp(i(\varphi_{31}+\varphi_{14}-\varphi_{11}))\\
&+&\frac{\beta}{\alpha}\exp(i(\varphi_{14}+\varphi_{31}+\varphi_{23}-\varphi_{12}-\varphi_{21}))\\
&\cdot&(d\sqrt{1-h^2}+hf(1-d^2)\sqrt{1-f^2}\exp(i(\varphi_{32}-\varphi_{22}))),\\
,\end{eqnarray*}
\begin{eqnarray*}u_{42}&=&-ab\sqrt{(1-d^2)(1-f^2)}\exp(i(\varphi_{12}+\varphi_{41}-\varphi_{11}))\\
&-&\frac{\alpha}{\beta}\sqrt(1-b^2)\exp(i(\varphi_{32}+\varphi_{41}-\varphi_{21}))(c\sqrt(1-h^2)\\
&-&bh\sqrt(1-c^2)\exp(i(\varphi_{23}-\varphi_{22})))\end{eqnarray*}
\begin{eqnarray*}
u_{43}&=&-ac\sqrt{(1-b^2)(1-d^2)(1-f^2)}\exp(i(\varphi_{13}+\varphi_{41}-\varphi_{11}))\\
&+&\frac{\alpha}{\beta}\exp(i(\varphi_{32}+\varphi_{13}+\varphi_{41}-\varphi_{12}-\varphi_{21}))\\
&\cdot&(b\sqrt{1-h^2}+ch(1-b^2)\sqrt{1-c^2}\exp(i(\varphi_{23}-\varphi_{22}))),\\
u_{44}&=&-a\sqrt{(1-b^2)(1-c^2)(1-d^2)(1-f^2)}\exp(i(\varphi_{14}+\varphi_{41}-\varphi_{11}))\\
&-&\alpha\beta h\exp(i(\varphi_{14}+\varphi_{41}+\varphi_{23}+\varphi_{32}-\varphi_{12}-\varphi_{21}-\varphi_{22})),
\end{eqnarray*}
where the parameters are the following
\begin{eqnarray*}\alpha = (f^2+d^2-f^2d^2)^{-1/2},\quad \beta = (b^2+c^2-b^2c^2)^{-1/2},\quad a,b,c,d,f,g\in[0,1].\end{eqnarray*}
\subsection{Appendix 2}
The elements of the matrix  (\ref{12_23}) are the following
\begin{eqnarray}\label{15_25_1}&&\rho_{12} = l_{1}u_{21}u_{11}^{*} + l_{2}u_{22}u_{12}^{*} + l_{3}u_{23}u_{13}^{*} + l_{4}u_{24}u_{14},\\\nonumber
&&\rho_{13} = l_{1}u_{11}u_{31}^{*} + l_{2}u_{12}u_{32}^{*} + l_{3}u_{13}u_{33}^{*} + l_{4}u_{14}u_{34},\\\nonumber
&&\rho_{14} = l_{1}u_{21}u_{31}^{*} + l_{2}u_{22}u_{32}^{*} + l_{3}u_{23}u_{33}^{*} + l_{4}u_{24}u_{34},\\\nonumber
&&\rho_{23} = l_{1}u_{11}u_{41}^{*} + l_{2}u_{12}u_{42}^{*} + l_{3}u_{13}u_{43}^{*} + l_{4}u_{14}u_{44},\\\nonumber
&&\rho_{24} = l_{1}u_{21}u_{41}^{*} + l_{2}u_{22}u_{42}^{*} + l_{3}u_{23}u_{43}^{*} + l_{4}u_{24}u_{44},\\\nonumber
&&\rho_{34} = l_{1}u_{41}u_{31}^{*} + l_{2}u_{42}u_{32}^{*} + l_{3}u_{43}u_{33}^{*} + l_{4}u_{44}u_{34},\\\nonumber
&&\rho_{11} = l_{1}|u_{11}|^2 + l_{2}|u_{12}|^2 + l_{3}|u_{13}|^2 + l_{4}|u_{14}|^2,\\\nonumber
&&\rho_{22} = l_{1}|u_{21}|^2 + l_{2}|u_{22}|^2 + l_{3}|u_{23}|^2 + l_{4}|u_{24}|^2,\\\nonumber
&&\rho_{33} = l_{1}|u_{31}|^2 + l_{2}|u_{32}|^2 + l_{3}|u_{33}|^2 + l_{4}|u_{34}|^2,\\\nonumber
&&\rho_{44} = l_{1}|u_{41}|^2 + l_{2}|u_{42}|^2 + l_{3}|u_{43}|^2 + l_{4}|u_{44}|^2.\nonumber
\end{eqnarray}
The elements of the matrix  (\ref{12_24}) are the following
\begin{eqnarray*}\rho_{11}&=&l_1|u_{11}|^2+l_3|u_{13}|^2,\quad\rho_{13}=l_1u_{11}u_{31}^{*}+l_3u_{13}u_{33}^{*},\\
\rho_{22}&=&l_4|u_{24}|^2+l_2|u_{22}|^2,\quad\rho_{24}=l_2u_{22}u_{42}^{*}+ l_4u_{24}u_{44}^{*},\\
\rho_{31}&=&l_1u_{11}^{*}u_{31}+ l_3 u_{13}^{*}u_{33},\quad\rho_{33}=l_3|u_{33}|^2+  l_1|u_{31}|^2,\\
\rho_{42}&=&l_2u_{22}^{*}u_{42}+ l_4u_{24}^{*}u_{44},\quad\rho_{44}=l_2|u_{42}|^2+ l_4|u_{44}|^2.
\end{eqnarray*}
The elements of the matrix  (\ref{12_25}) are the following
\begin{eqnarray*}\rho_{11}&=&l_1|u_{11}|^2+l_2|u_{12}|^2,\quad\rho_{12}=l_1u_{11}u_{21}^{*}+l_2u_{12}u_{22}^{*},\\
\rho_{21}&=&l_1u_{21}u_{11}^{*}+l_2u_{22}u_{12}^{*},\quad\rho_{22}=l_1|u_{21}|^2+ l_2|u_{22}|^2,\\
\rho_{33}&=&l_3|u_{33}|^2+  l_4|u_{34}|^2,\quad\rho_{34}=l_3u_{33}u_{43}^{*}+  l_4u_{34}u_{44}^{*},\\
\rho_{43}&=&l_4u_{44}u_{34}^{*}+ l_3u_{33}^{*}u_{43},\quad\rho_{44}=l_3|u_{43}|^2+ l_4|u_{44}|^2.
\end{eqnarray*}
The elements of the matrix  (\ref{15_30}) are the following
\begin{eqnarray*}\rho_{11}&=&l_1|u_{11}|^2+l_4|u_{14}|^2,\quad\rho_{14}=l_1u_{11}u_{41}^{*}+l_4u_{14}u_{44}^{*},\\
\rho_{22}&=&l_3|u_{23}|^2+l_2|u_{22}|^2,\quad\rho_{23}=l_2u_{22}u_{32}^{*}+ l_3u_{23}u_{33}^{*},\\
\rho_{32}&=&l_2u_{22}^{*}u_{32}+ l_3 u_{23}^{*}u_{33},\quad\rho_{33}=l_3|u_{33}|^2+  l_2|u_{32}|^2,\\
\rho_{41}&=&l_1u_{11}^{*}u_{41}+ l_4u_{14}^{*}u_{44},\quad\rho_{44}=l_1|u_{41}|^2+ l_4|u_{44}|^2.
\end{eqnarray*}
\end{document}